\documentclass[prl,twocolumn,superscriptaddress,showpacs,floatfix]{revtex4}
\usepackage{epsfig}
\usepackage{dcolumn}  
\usepackage{bm}       
\usepackage{amsfonts, amssymb}
\usepackage{url}
\newcommand{\tc}{,~}
\newcommand{\gevsq}{GeV$^2$}

\newcommand{\cma}{$\theta_{cm}$}
\newcommand{\invs}{$s$}
\newcommand{\invt}{$t$}

\newcommand{\KLL}{$K_{_{LL}}$}
\newcommand{\rea}{H($\vec \gamma, \vec p \gamma)$}
\newcommand{\PRD}{Phys. Rev. D}
\newcommand{\PRC}{Phys. Rev. C}
\newcommand{\PRL}{ Phys. Rev. Lett.}

\newcommand{\NP}{Nucl. Phys.}
\newcommand{\NIM}{Nucl. Instr. Meth.}
\newcommand{\etal}{{\em et al.}}
\begin{document}

\preprint{}

\pacs{13.60.Fz,24.85.+p}

\title{Compton Scattering Cross Section on
the Proton at High Momentum Transfer}

\author{A.~Danagoulian}
\affiliation{\mbox{University of Illinois, Urbana-Champaign, IL 61801}}
\author{V.~H.~Mamyan}
\affiliation{\mbox{Yerevan Physics Institute, Yerevan 375036,
Armenia}} \affiliation{\mbox{Thomas Jefferson National Accelerator
Facility\tc Newport News, VA 23606}}
\author{M.~Roedelbronn}
\affiliation{\mbox{University of Illinois, Urbana-Champaign, IL 61801}}
\author{K.~A.~Aniol}
\affiliation{\mbox{California State University Los Angeles\tc
Los Angeles\tc CA 90032}}
\author{J.~R.~M.~Annand}
\affiliation{\mbox{University of Glasgow, Glasgow G12 8QQ, Scotland, U.K.}}
\author{P.~Y.~Bertin}
\affiliation{\mbox{Universit\'{e} Blaise Pascal/IN2P3\tc
F-63177 Aubi\`{e}re, France}}
\author{L.~Bimbot}
\affiliation{\mbox{IPN Orsay B.P. n$^\circ$1 F-91406, Orsay, France}}
\author{P.~Bosted}
\affiliation{\mbox{University of Massachusetts, Amherst, MA 01003}}
\author{J.~R.~Calarco}
\affiliation{\mbox{University of New Hampshire, Durham, NH 03824}}
\author{ A.~Camsonne}
\affiliation{\mbox{Universit\'{e} Blaise Pascal/IN2P3\tc
F-63177 Aubi\`{e}re, France}}
\author{C.~C.~Chang}
\affiliation{\mbox{University of Maryland, College Park, MD 20742}}
\author{T.-H.~Chang}
\affiliation{\mbox{University of Illinois, Urbana-Champaign, IL 61801}}
\author{J.-P.~Chen}
\affiliation{\mbox{Thomas Jefferson National Accelerator Facility\tc
Newport News, VA 23606}}
\author{Seonho~Choi}
\affiliation{\mbox{Temple University, Philadelphia, PA 19122}}
\author{E.~Chudakov}
\affiliation{\mbox{Thomas Jefferson National Accelerator Facility\tc
Newport News, VA 23606}}
\author{P.~Degtyarenko}
\affiliation{\mbox{Thomas Jefferson National Accelerator Facility\tc
Newport News, VA 23606}}
\author{C.~W.~de~Jager}
\affiliation{\mbox{Thomas Jefferson National Accelerator Facility\tc
Newport News, VA 23606}}
\author{A.~Deur}
\affiliation{\mbox{University of Virginia, Charlottesville, VA 22901}}
\author{D.~Dutta}
\affiliation{\mbox{Duke University and TUNL, Durham, NC 27708}}
\author{K.~Egiyan}\altaffiliation{Deceased}
\affiliation{\mbox{Yerevan Physics Institute, Yerevan 375036, Armenia}}
\author{H.~Gao}
\affiliation{\mbox{Duke University and TUNL, Durham, NC 27708}}
\author{F.~Garibaldi}
\affiliation{\mbox{INFN, Sezione di Sanit\'{a}
and Institute Superiore di Sanit\'{a} \tc I-00161 Rome\tc Italy}}
\author{O.~Gayou}
\affiliation{\mbox{College of William and Mary, Williamsburg, VA 23187}}
\author{R.~Gilman}
\affiliation{\mbox{Thomas Jefferson National Accelerator Facility\tc
Newport News, VA 23606}}
\affiliation{\mbox{Rutgers, The State University of New Jersey\tc
Piscataway, NJ 08854}}
\author{A.~Glamazdin}
\affiliation{\mbox{Kharkov Institute of Physics and Technology\tc
Kharkov 61108, Ukraine}}
\author{C.~Glashausser}
\affiliation{\mbox{Rutgers, The State University of New Jersey\tc
Piscataway, NJ 08854}}
\author{J.~Gomez}
\affiliation{\mbox{Thomas Jefferson National Accelerator Facility\tc
Newport News, VA 23606}}
\author{D.~J.~Hamilton}
\affiliation{\mbox{University of Glasgow, Glasgow G12 8QQ, Scotland, U.K.}}
\author{J.-O.~Hansen}
\affiliation{\mbox{Thomas Jefferson National Accelerator Facility\tc
Newport News, VA 23606}}
\author{D.~Hayes}
\affiliation{\mbox{Old Dominion University, Norfolk, VA 23529}}
\author{D.~W.~Higinbotham}
\affiliation{\mbox{Thomas Jefferson National Accelerator Facility\tc
Newport News, VA 23606}}
\author{W.~Hinton}
\affiliation{\mbox{Old Dominion University, Norfolk, VA 23529}}
\author{T.~Horn}
\affiliation{\mbox{University of Maryland, College Park, MD 20742}}
\author{C.~Howell}
\affiliation{\mbox{Duke University and TUNL, Durham, NC 27708}}
\author{T.~Hunyady}
\affiliation{\mbox{Old Dominion University, Norfolk, VA 23529}}
\author{C.~E.~Hyde-Wright}
\affiliation{\mbox{Old Dominion University, Norfolk, VA 23529}}
\author{X.~Jiang}
\affiliation{\mbox{Rutgers, The State University of New Jersey\tc
Piscataway, NJ 08854}}
\author{M.~K.~Jones}
\affiliation{\mbox{Thomas Jefferson National Accelerator Facility\tc
Newport News, VA 23606}}
\author{M.~Khandaker}
\affiliation{\mbox{Norfolk State University, Norfolk, VA 23504}}
\author{A.~Ketikyan}
\affiliation{\mbox{Yerevan Physics Institute, Yerevan 375036, Armenia}}
\author{V.~Koubarovski}
\affiliation{\mbox{Rensselaer Polytechnic Institute, Troy, NY 12180}}
\author{K.~Kramer}
\affiliation{\mbox{College of William and Mary, Williamsburg, VA 23187}}
\author{G.~Kumbartzki}
\affiliation{\mbox{Rutgers, The State University of New Jersey\tc
Piscataway, NJ 08854}}
\author{G.~Laveissi\`ere}
\affiliation{\mbox{Universit\'{e} Blaise Pascal/IN2P3\tc
F-63177 Aubi\`{e}re, France}}
\author{J.~LeRose}
\affiliation{\mbox{Thomas Jefferson National Accelerator Facility\tc
Newport News, VA 23606}}
\author{R.~A.~Lindgren}
\affiliation{\mbox{University of Virginia, Charlottesville, VA 22901}}
\author{D.~J.~Margaziotis}
\affiliation{\mbox{California State University Los Angeles\tc Los Angeles\tc CA 90032}}
\author{P.~ Markowitz}
\affiliation{\mbox{Florida International University, Miami, FL 33199}}
\author{K.~McCormick}
\affiliation{\mbox{Old Dominion University, Norfolk, VA 23529}}
\author{Z.-E.~Meziani}
\affiliation{\mbox{Temple University, Philadelphia, PA 19122}}
\author{R.~Michaels}
\affiliation{\mbox{Thomas Jefferson National Accelerator Facility\tc
Newport News, VA 23606}}
\author{P.~Moussiegt}
\affiliation{\mbox{Institut des Sciences Nucleiares\tc CNRS-IN2P3\tc
F-38016 Grenoble, France}}
\author{S.~Nanda}
\affiliation{\mbox{Thomas Jefferson National Accelerator Facility\tc
Newport News, VA 23606}}
\author{A.~M.~Nathan}
\affiliation{\mbox{University of Illinois, Urbana-Champaign, IL 61801}}
\author{D.~M.~Nikolenko}
\affiliation{\mbox{Budker Institute for Nuclear Physics\tc
Novosibirsk 630090, Russia}}
\author{V.~Nelyubin}
\affiliation{\mbox{ St.~Petersburg Nuclear Physics Institute\tc
Gatchina, 188350, Russia}}
\author{B.~E.~Norum}
\affiliation{\mbox{University of Virginia, Charlottesville, VA 22901}}
\author{K.~Paschke}
\affiliation{\mbox{University of Massachusetts, Amherst, MA 01003}}
\author{L.~Pentchev}
\affiliation{\mbox{College of William and Mary, Williamsburg, VA 23187}}
\author{C.~F.~Perdrisat}
\affiliation{\mbox{College of William and Mary, Williamsburg, VA 23187}}
\author{E.~Piasetzky}
\affiliation{\mbox{Tel Aviv University, Tel Aviv 69978, Israel}}
\author{R.~Pomatsalyuk}
\affiliation{\mbox{Kharkov Institute of Physics and Technology\tc
Kharkov 61108, Ukraine}}
\author{V.~A.~Punjabi}
\affiliation{\mbox{Norfolk State University, Norfolk, VA 23504}}
\author{I.~Rachek}
\affiliation{\mbox{Budker Institute for Nuclear Physics\tc
Novosibirsk 630090, Russia}}
\author{A.~Radyushkin}
\affiliation{\mbox{Thomas Jefferson National Accelerator Facility\tc
Newport News, VA 23606}}
\affiliation{\mbox{Old Dominion University, Norfolk, VA 23529}}
\author{B.~Reitz}
\affiliation{\mbox{Thomas Jefferson National Accelerator Facility\tc
Newport News, VA 23606}}
\author{R.~Roche}
\affiliation{\mbox{Florida State University, Tallahassee, FL 32306}}
\author{G.~Ron}
\affiliation{\mbox{Tel Aviv University, Tel Aviv 69978, Israel}}
\author{F.~Sabati\'e}
\affiliation{\mbox{Old Dominion University, Norfolk, VA 23529}}
\author{A.~Saha}
\affiliation{\mbox{Thomas Jefferson National Accelerator Facility\tc
Newport News, VA 23606}}
\author{N.~Savvinov}
\affiliation{\mbox{University of Maryland, College Park, MD 20742}}
\author{A.~Shahinyan}
\affiliation{\mbox{Yerevan Physics Institute, Yerevan 375036, Armenia}}
\author{Y.~Shestakov}
 \affiliation{\mbox{Budker Institute for Nuclear Physics\tc Novosibirsk
630090, Russia}}
\author{S.~\v{S}irca}
\affiliation{\mbox{Massachusetts Institute of Technology, Cambridge, MA 02139}}
\author{K.~Slifer}
\affiliation{\mbox{Temple University, Philadelphia, PA 19122}}
\author{P.~Solvignon}
\affiliation{\mbox{Temple University, Philadelphia, PA 19122}}
\author{P.~Stoler}
\affiliation{\mbox{Rensselaer Polytechnic Institute, Troy, NY 12180}}
\author{S.~Tajima}
\affiliation{\mbox{Duke University and TUNL, Durham, NC 27708}}
\author{V.~Sulkosky}
\affiliation{\mbox{College of William and Mary, Williamsburg, VA 23187}}
\author{L.~Todor}
\affiliation{\mbox{Old Dominion University, Norfolk, VA 23529}}
\author{B.~Vlahovic}
\affiliation{\mbox{North Carolina Central University, Durham, NC 27707}}
\author{L.~B.~Weinstein}
\affiliation{\mbox{Old Dominion University, Norfolk, VA 23529}}
\author{K.~Wang}
\affiliation{\mbox{University of Virginia, Charlottesville, VA 22901}}
\author{B.~Wojtsekhowski}
\affiliation{\mbox{Thomas Jefferson National Accelerator Facility\tc
Newport News, VA 23606}}
\author{H.~Voskanyan}
\affiliation{\mbox{Yerevan Physics Institute, Yerevan 375036, Armenia}}
\author{H.~Xiang}
\affiliation{\mbox{Massachusetts Institute of Technology, Cambridge, MA 02139}}
\author{X.~Zheng}
\affiliation{\mbox{Massachusetts Institute of Technology, Cambridge, MA 02139}}
\author{L.~Zhu}
\affiliation{\mbox{Massachusetts Institute of Technology, Cambridge, MA 02139}}

\collaboration{The Jefferson Lab Hall A Collaboration}
\noaffiliation{}

\date{\today}

\begin{abstract}
Cross-section values for Compton scattering on the proton were
measured at 25 kinematic settings over the range \mbox{\invs~=~5
-- 11} and \mbox{-\invt~=~2 -- 7~\gevsq} with statistical accuracy
of a few percent. The scaling power for the $s$-dependence of the
cross section at fixed center of mass angle was found to be
8.0$\pm$0.2, strongly inconsistent with the prediction of
perturbative QCD. The observed cross-section values are in fair
agreement with the calculations using the handbag mechanism, in
which the external photons couple to a single quark.
\end{abstract}

\pacs{13.60.Fz,24.85.+p}
\maketitle

Compton scattering in its various forms provides a unique tool for
studying many aspects of hadronic structure by probing it with two
electromagnetic currents. For real Compton scattering (RCS) in the
hard scattering regime, where all Mandelstam variables $s$, $-t$,
and $-u$ are larger than the $\Lambda^2_{_{QCD}}$ scale, the
short-distance dominance is secured by the presence of a large
momentum transfer. In this regime, RCS probes the fundamental
quark-gluon degrees of freedom of quantum chromodynamics (QCD),
providing important information for the tomographic imaging of the
nucleon.

The only data for RCS in the hard scattering regime were obtained
25~years ago by the pioneering Cornell experiment~\cite{sh79}. The
cross section $d\sigma/dt$ at fixed \cma~was found to scale with
$1/s^n$ with $n\approx6$, exactly as predicted by perturbative
QCD~\cite{br73}, in which the reaction is mediated by the exchange
of two hard gluons~\cite{kr91}.  Nevertheless, the experimental
cross section was at least 10 times larger than those predicted by
perturbative QCD.  More recently, calculations of RCS have been
performed within a handbag dominance model~\cite{ra98,di99}, in
which the external photons couple to a single quark, which couples
to the spectator particles through generalized parton distributions
(GPDs)~\cite{di05}. These calculations are rather close to the
Cornell cross section data. The uncertainty in applicability of
perturbative QCD and the possible dominance of the handbag mechanism
were reinforced by a recent measurement of the longitudinal
polarization transfer parameter \KLL~in the reaction
\rea~\cite{ha05}, which is in fair agreement with the handbag
prediction~\cite{hu02}
and in unambiguous disagreement with the perturbative QCD
prediction~\cite{kr91}. The present experiment was designed to
test more stringently the reaction mechanism by improving the
statistical precision and extending the kinematic range of the
Cornell data.  These new measurements, with much improved accuracy
in the scaling parameter $n$, allow unambiguous conclusions about
the applicability of perturbative QCD.

\begin{figure}[tb]
\epsfig{file=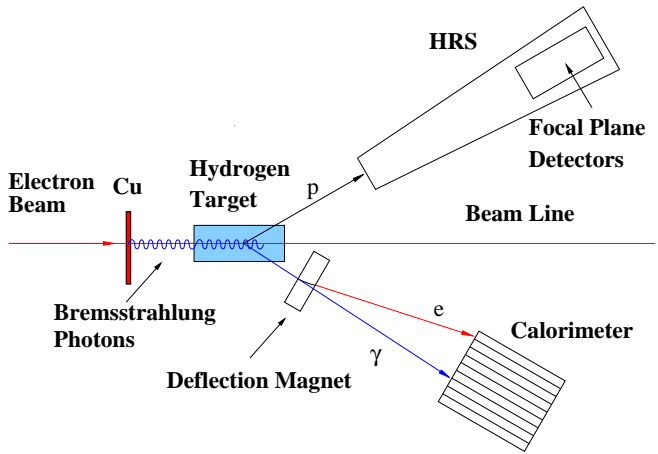,angle=0,width=\linewidth}
\caption{Schematic layout of the present experiment.}
\label{fig:setup}
\end{figure}

The experiment, shown schematically in Fig.~\ref{fig:setup}, was
performed in Hall A of Jefferson Lab, with basic instrumentation
described in Ref.~\cite{al04}. A 100\% duty-factor electron beam
with current up to 40~$\mu$A and energy 2.34, 3.48, 4.62, or 5.75~GeV
was incident on a 0.81-mm thick Cu radiator. The mixed beam of
electrons and bremsstrahlung photons was incident on a 15-cm
liquid H$_2$ target, located 10~cm downstream of the radiator, with
a photon flux of up to $2 \times 10^{13}$ equivalent quanta/s.
The scattered photon was detected in a calorimeter consisting of
704 lead-glass blocks ($4\times 4\times 40$~cm$^3$) placed 5-18~m
from the target and subtending a solid angle of 30-60~msr, with a
typical position resolution of 1~cm and energy resolution
$\sigma_E/E \,=\, 5 - 10$\%. The associated recoil proton was
detected in one of the Hall A High Resolution Spectrometers (HRS),
with a solid angle of 6.5~msr, momentum acceptance of $\pm$4.5\%,
relative momentum resolution of $2.5 \times 10^{-4}$, and angular
resolution of 2.4~mrad, the latter limited principally by
scattering in the target. The central momentum of the HRS was set
to detect protons corresponding to incident photons with mean
energies approximately 90\% of the electron beam energy. The
trigger was formed from a coincidence between a signal from a
scintillator counter in the HRS focal plane and a signal in the
calorimeter corresponding to an energy deposition greater than
half the expected photon energy from the RCS process.

\begin{figure}[htb]
\epsfig{file=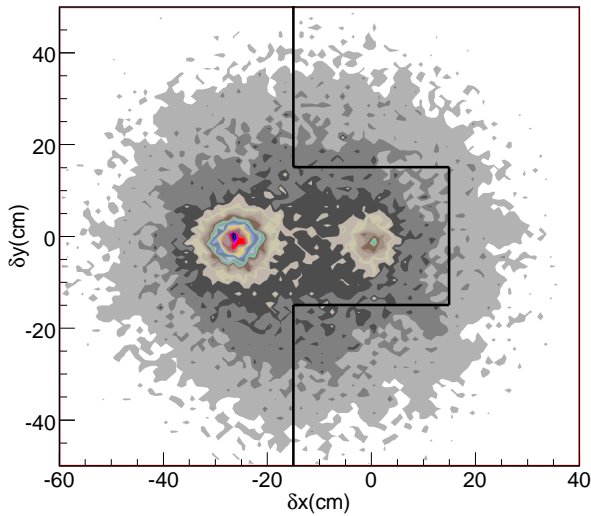,angle=00,width=\linewidth}
\caption{Distribution of photon-proton coincidence events in
$\delta x$-$\delta y$ space, as defined in the text, for the
measurement at $ s = 6.79, t = - 3.04$~\gevsq.  The peaks near
$\delta x=\delta y=0$ and $\delta x=-25$~cm, $\delta y=0$ originate
from the RCS and $ep$ events, respectively. The continuum events are
photons from $\pi^\circ$ photoproduction. The events to the right
of the solid line are used to normalize the $\pi^\circ$ events
yield in the Monte Carlo simulation.} \label{fig:dxdy}
\end{figure}

Potential RCS events were within a $\sim30$~ns coincidence time
window and were selected based on the kinematic correlation between the
scattered photon and the recoil proton.
The excellent HRS optics was used to reconstruct the momentum
vector and reaction vertex of the recoil proton, to determine the
energy of the incident photon, and to calculate the expected
direction of an RCS photon.  The quantities $\delta x$ and $\delta
y$ are the difference of horizontal and vertical coordinates,
respectively, between the expected and measured location of the
detected photon on the front face of the calorimeter.
An example of the distribution of events in the $\delta x$-$\delta y$
plane is shown in Fig.~\ref{fig:dxdy}.
The RCS events, which are in the peak at $\delta x=\delta y=0$,
lie upon a continuum background primarily from
the $\gamma p\rightarrow\pi^\circ p$ reaction, with the
subsequent decay $\pi^\circ\rightarrow\gamma\gamma$.
An additional background is due to electrons that lose energy in
the radiator and subsequently undergo $ep$ elastic scattering,
which is kinematically indistinguishable from RCS.
A magnet between the target and the calorimeter (see Fig.~\ref{fig:setup})
deflects these electrons horizontally, so their coordinates
on the front face of the calorimeter are shifted by 20-30~cm relative to
undeflected RCS photons.
These $ep$ events are clearly separated from the RCS events,
as shown in Fig.~\ref{fig:dxdy}.
\begin{figure}[htb]
\epsfig{file=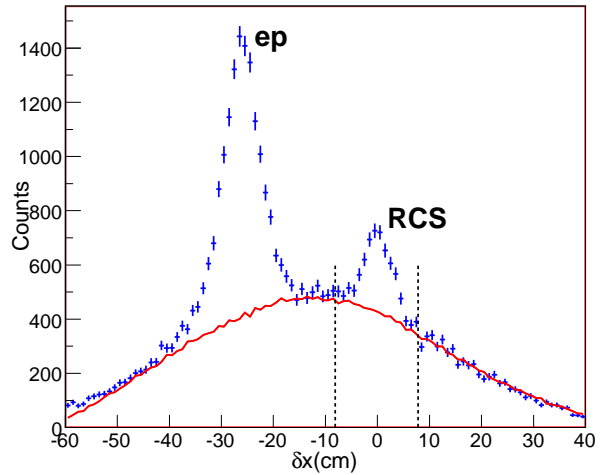,angle=0,width=\linewidth} \caption{The
$\delta x$-distribution for a coplanarity cut $|\delta y|
\leqslant 15$~cm, with the RCS and $ep$ peaks indicated. The curve
is a distribution of the continuum $\pi^\circ$ events. The
vertical dashed lines show the cuts used to calculate the number
of RCS events.} \label{fig:dx}
\end{figure}
By making the coplanarity cut $|\delta y| \leqslant 15$~cm, then
projecting onto the $\delta x$ axis, one obtains the $\delta
x$-distribution shown in Fig.~\ref{fig:dx}. The curve is a
calculation of the $\pi^\circ$ continuum background, which is
determined by two methods.

In the first method, a Monte Carlo simulation of the experiment
was used to determine the acceptance of the combined
HRS-calorimeter system in the variables of the incident photon
energy and the momentum transfer, and to determine the shape of
$\pi^\circ$ contribution in Fig.~\ref{fig:dx}~\cite{areg}. The
simulation utilized a thick-target bremsstrahlung code to
calculate the incident photon spectrum~\cite{mo73}; an event
generator for the RCS, $\gamma p \rightarrow \pi^\circ p$, and
$e\,p \rightarrow e\,p$ reactions; and the SIMC code~\cite{simc}
to track recoil protons through the HRS. The acceptance simulation
and analysis procedures were checked by using elastic electron
scattering data from dedicated H$(e,e^\prime p)$ runs with the Cu
radiator removed. It was verified that the simulation correctly
accounts for the distribution of proton recoil events in momentum,
angles, and reaction vertex across the acceptance of the
HRS~\cite{da05} and that the data from the present experimental
setup reproduce to better than 3\% the known $ep$ elastic
scattering cross section~\cite{ar04}. To determine the $\pi^\circ$
background, the simulated distribution of $\pi^\circ$ events is
normalized to the number of actual events in regions of $\delta
x$-$\delta y$ space that are free of RCS and $ep$ events (see
Fig.~\ref{fig:dxdy}), then used to calculate the curve in
Fig.~\ref{fig:dx}. The tight kinematic constraints of our
coincidence geometry preclude contributions from heavier mesons,
such as $\eta$'s. Subtracting the curve from the data, then
integrating over the region of $\delta x$ shown in
Fig.~\ref{fig:dx}, the ``raw'' RCS cross section is determined. As
a byproduct of this analysis, we have obtained cross sections for
the $p(\gamma,\pi^\circ)p$ reaction, which will be presented in a
separate publication.

The second method of analysis uses only a central \mbox{($\sim35$\%)}
portion of the the calorimeter front face to guarantee that the
combined acceptance of the experiment is defined by the photon arm
acceptance~\cite{vahe}.  Events were selected in a narrow energy
range, 100-200~MeV, in which the incident photon spectrum had the
expected $1/E_\gamma$ shape.
The shape of the $\delta x$-distribution of the $\pi^\circ$ events
in Fig.~\ref{fig:dx} was obtained by interpolation of a polynomial
fit to the event distribution in the region $15 \leqslant |\delta
y| \leqslant 30$~cm of Fig.~\ref{fig:dxdy}. The total $\pi^\circ$
yield was normalized to the regions $\delta x\geqslant 10$~cm and
 $|\delta x + 10$~cm$| \leqslant 3$~cm in Fig.~\ref{fig:dx}.
As with the Monte Carlo method, subtracting the $\pi^0$ background
and integrating over $\delta x$ obtains the raw RCS cross section,
with reduced statistical precision.  The two methods agreed to
approximately  5\%.  The Monte Carlo technique was used for the
final cross-section values while the reduced acceptance technique was
used to estimate the contribution of acceptance and $\pi^0$
background to the systematic error.

Two additional corrections were applied to obtain the RCS cross
section. The first correction deals with the kinematically
correlated $p\gamma$ background events from the $ep\gamma$
process, in which an elastically scattered electron emits a hard
photon due to internal and external radiation in the target and
surrounding material and the photon (but not the electron) is
detected in the calorimeter. These background events fall in the
$\delta x$=0 peak in Fig.~\ref{fig:dx} and are distinguishable
from RCS events by $E_{calo}$, the photon energy measured in the
calorimeter.  To determine the background from these events, a
semi-empirical technique was used.

First the shape of the photon energy spectrum, in which an
elastically scattered electron radiates a hard photon in the
material between the reaction vertex and the deflection magnet,
was found from the Monte Carlo simulation, which includes all
details of the experimental setup. Then the shape was normalized
to fit the observed distribution in the $E_{calo}$ spectrum below
the RCS peak. The resulting background was subtracted and the peak
integrated over a $\pm 3\sigma$ region to obtain the RCS events. A
similar procedure was applied to the electron scattering data
taken with the radiator removed to obtain another normalization.
The $ep\gamma$/RCS ratio ranges from $<0.01$ at backward angles to
as much as 0.90 at forward angles. Nevertheless, the two
normalizations result in RCS cross sections that agree to within a
statistical accuracy of 7\% in the worst case but more typically
to within 2\%.  This procedure was cross-checked against a direct
calculation of the background, using the peaking
approximation~\cite{rolf} to estimate the internal radiation
contribution and found to be in excellent agreement.

The second correction is due to quasi-real photons from the
H$(e,p\gamma)e^\prime$ reaction and is taken into account in the
calculation of the incident photon flux. The reaction is simulated
with our Monte Carlo, using the spectrum of quasi-real photons
calculated according to the method of Ref.~\cite{bu75}.
Although the scattered electron is not detected,
the kinematic cuts on the HRS and calorimeter, particularly
the $\delta x$ and $\delta y$ cuts,
place stringent constraints on the virtuality of the photon.
We find that the quasi-real photons have a mean \mbox{$Q^2 \,=\,
0.14 \times 10^{-3}$~\gevsq} and contribute in the range 11-15\% to
the total incident photon flux, depending on the kinematic point.

The resulting RCS cross-section values and statistical uncertainties are
summarized in Table~\ref{tab:results}.  The systematic
uncertainties have correlated contributions from the acceptance
(5\%) and the real and virtual bremsstrahlung flux uncertainty
(3\%) and typical point-to-point contributions from the
$\pi^\circ$ subtraction (3\%) and the $ep\gamma$ background
subtraction (2\%).

\begin{table}[htb]
\caption{Cross section of proton Compton scattering. The mean
values of invariants \invs, \invt~and their standard deviation are
in \gevsq. The scattering angle in the center-of-mass system and
its standard deviation are in degrees. The bin widths of all
quantities are the total spread in values over the acceptance of
the detectors. The cross section ($d \sigma /dt$) and its
statistical error are in nb/\gevsq.}
\begin{ruledtabular}
\renewcommand{\arraystretch}{1.5}
\begin{tabular}{lccccccc}
 $s$  & $\Delta s$ & $-t$   & $\Delta t$  &
$\theta_{cm}$  & $\Delta\theta$  &  $d \sigma /dt$ &
          $\Delta d \sigma /dt$ \\
\hline
 4.82  & 0.56  & 1.65 & 0.05 &  90.0 & 1.0  & 6.37   & 0.18\\
 4.82  & 0.56  & 2.01 & 0.06 & 104.4 & 1.3  & 4.59   & 0.13\\
 4.82  & 0.56  & 2.60 & 0.08 & 127.9 & 1.8  & 2.18   & 0.05\\
 6.79  & 0.56  & 1.96 & 0.05 &  76.3 & 0.8  & 0.815  & 0.040 \\
 6.79  & 0.56  & 2.54 & 0.06 &  89.2 & 1.0  & 0.251  & 0.027\\
 6.79  & 0.56  & 3.04 & 0.07 &  100.5 & 1.1 & 0.226  & 0.018 \\
 6.79  & 0.56  & 3.70 & 0.08 & 115.9 & 1.3  & 0.282  & 0.009 \\
 6.79  & 0.56  & 4.03 & 0.08 & 124.5 & 1.3  & 0.291  & 0.009 \\
 6.79  & 0.56  & 4.35 & 0.09 & 133.7 & 1.4  & 0.304  & 0.011  \\
 8.90 & 0.84  & 2.03 & 0.05 &  64.0 & 0.8   & 0.3970 & 0.0211  \\
 8.90 & 0.84  & 2.57 & 0.06 &  73.2 & 0.8   & 0.1109 & 0.0078  \\
 8.90 & 0.84  & 3.09 & 0.07 &  81.6 & 0.9   & 0.0619 & 0.0055   \\
 8.90 & 0.84  & 3.68 & 0.08 &  91.0 & 1.1   & 0.0348 & 0.0029   \\
 8.90 & 0.84  & 4.38 & 0.09 & 102.3 & 1.1   & 0.0257 & 0.0028   \\
 8.90 & 0.84  & 5.03 & 0.09 & 113.1 & 1.2   & 0.0320 & 0.0035   \\
 8.90 & 0.84  & 5.48 & 0.10 & 121.0 & 1.2   & 0.0477 & 0.0031   \\
 8.90 & 0.84  & 5.92 & 0.10 & 129.8 & 1.2   & 0.0641 & 0.0042   \\
10.92 & 0.94  & 2.61 & 0.08 &  65.3 & 0.9   & 0.0702 & 0.0063   \\
10.92 & 0.94  & 3.18 & 0.09 &  71.9 & 0.9   & 0.0317 & 0.0047   \\
10.92 & 0.94  & 3.73 & 0.10 &  79.0 & 1.0   & 0.0156 & 0.0026   \\
10.92 & 0.94  & 4.41 & 0.12 &  87.5 & 1.1   & 0.0095 & 0.0011   \\
10.92 & 0.94  & 5.03 & 0.14 &  94.1 & 1.2   & 0.0071 & 0.0007   \\
10.92 & 0.94  & 5.44 & 0.14 &  100.3 & 1.3  & 0.0058 & 0.0009   \\
10.92 & 0.94  & 5.93 & 0.16 & 106.6 & 1.3   & 0.0046 & 0.0006   \\
10.92 & 0.94  & 6.46 & 0.19 & 113.6 & 2.1   & 0.0056 & 0.0007
\end{tabular}
\end{ruledtabular}
\label{tab:results}
\end{table}

\begin{figure}[htb]
\epsfig{file=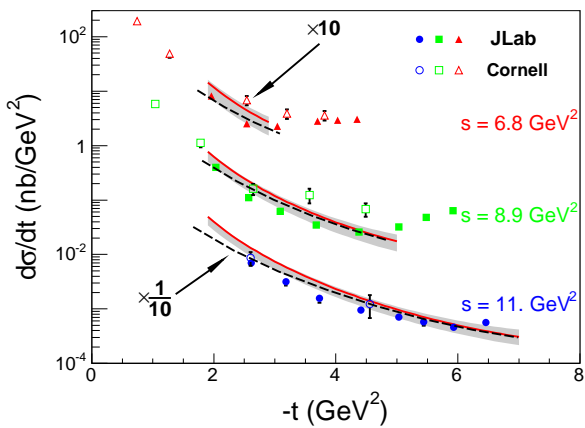,angle=0,width=\linewidth}
\caption{ Cross section
of RCS process vs. transfer momentum $t$ at three values of $s$.
Full points and open points are data from the present and Cornell
experiments~\cite{sh79}, respectively.} \label{fig:crsec}
\end{figure}

The cross-section data are presented in Fig.~\ref{fig:crsec} along
with the previous Cornell data~\cite{sh79}, which have been scaled
to the $s$ values of the present experiment using the scaling
power $n=8$, as discussed below, and plotted at the $-t$ value
corresponding to the original $\theta_{cm}$.
The curves are theoretical predictions calculated with the handbag
diagram. The solid curves are calculations using the GPDs
approach~\cite{hu02}, in which a photon-parton subprocess is
calculated to next-to-leading order in $\alpha_{s}$ and a model of
the GPDs is based on the known parton distribution functions and
the nucleon electromagnetic form factors.
The widths of the shaded areas indicate the uncertainties due to the
mass uncertainties in the hard subprocess~\cite{di03b}.
The dashed curves are also based on the handbag diagram~\cite{mi04},
using the constituent quark model to calculate the hard subprocess
and quark wave functions adjusted to fit existing data for the
nucleon electromagnetic form factors.
Both sets of curves cover a limited range in $-t$ because
the calculations based on the handbag mechanism are valid only
for $s, -t, -u$ larger than approximately 2.5~\gevsq.
Over that range they are in good agreement with the data.

\begin{figure}[htb]
\epsfig{file= 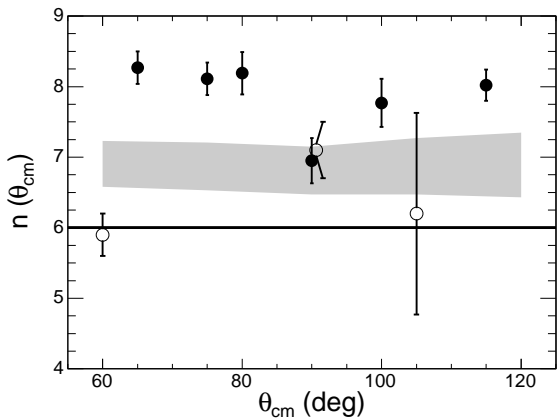,angle=0,width=\linewidth}
\caption{Scaling of the RCS cross section at fixed \cma. Open
points are results from Cornell experiment~\cite{sh79}. Closed
points are results from the present experiment. The line at $n=6$
is the prediction of asymptotic perturbative QCD, while the shaded
area shows the fit range obtained from the cross sections of
GPDs-based handbag calculation~\cite{hu02}.} \label{fig:ntheta}
\end{figure}
It is interesting to examine the scaling of the cross sections
with $s$ at fixed \cma, where the perturbative QCD mechanism
predicts $d\sigma/dt=f(\theta_{cm})/s^{n}$ with $n=6$~\cite{br73}.
The scaling power $n({\theta}_{cm})$ was extracted from the
present data by using results from the three largest values of
\mbox{$s$ = 6.79, 8.90, and 10.92~\gevsq}.  A cubic spline interpolation was
applied to the angular distribution for each $s$ to determine the
cross section at fixed angles.  The values of $n({\theta}_{cm})$
are plotted in Fig.~\ref{fig:ntheta} along with points from the
Cornell experiment.  The present experimental points imply a mean
value $n=8.0 \pm 0.2$, unequivocally demonstrating that the
perturbative QCD mechanism is not dominant in the presently
accessible kinematic range.  The power obtained from a fit to
GPDs-based handbag cross sections~\cite{hu02} are shown as the
dashed lines for two different assumptions about the masses in the
hard subprocess~\cite{di03b}.  The present data should help refine
the model used for the GPDs.

In summary, the RCS cross section from the proton was measured in
range $s = 5 - 11 $~\gevsq~at large momentum transfer. Calculations
based on the handbag diagram are in good agreement with experimental
data, suggesting that the reaction mechanism in the few GeV energy
range is dominantly one in which the external photons couple to a
single quark.  The fixed-\cma\ scaling power is considerably larger
than that predicted by perturbative QCD.

We thank P.~Kroll, J.~M.~Laget, and G.~Miller for productive
discussions, and acknowledge the Jefferson Lab staff for their
outstanding contributions. This work was supported  the US
Department of Energy  under contract DE-AC05-84ER40150,
Modification No. M175, under which  the Southeastern Universities
Research Association (SURA) operates the Thomas Jefferson National
Accelerator Facility. We acknowledge additional grants from the
U.S. National Science Foundation, the UK Engineering and Physical
Science Research Council, the Italian INFN, the French CNRS and
CEA, and the Israel Science Foundation.

\end{document}